\documentclass[aps,pra,twocolumn,letter]{revtex4-1}
\usepackage[english]{babel}
\usepackage{subeqnarray}
\usepackage{hyperref}
\usepackage{graphicx}
\usepackage{mathrsfs}
\usepackage{bm}
\usepackage{amssymb}
\usepackage{amsmath}
\usepackage{subfigure}

\newcommand{\vv}[1]{\mathbf{#1}}
\newcommand{\bra}[1]{\langle#1|}
\newcommand{\ket}[1]{|#1\rangle}
\newcommand{\pro}[2]{\ket{#1}\bra{#2}}
\newcommand{\eq}[1]{(\ref{eq:#1})}
\newcommand{\fig}[1]{Fig.~\ref{fig:#1}}
\newcommand{\abs}[1]{\left|#1\right|}
\newcommand{\real}[1]{\mathrm{Re}[#1]}

\begin{document}

\title{Dephasing dynamics of Rydberg atom spin waves}
\date{\today}
\author{F.~Bariani}
\author{Paul~M.~Goldbart}
\author{T.A.B.~Kennedy}
\affiliation{School of Physics, 837 State Street, Georgia Institute of Technology, Atlanta, GA, 30332-0430, USA}

% Pacs: 32.80.Ee, 32.80.Qk, 42.50.Dv, 42.50.Ar

\begin{abstract}
A theory of Rydberg atom interactions is used to derive analytical forms for the spin wave pair correlation function in laser-excited cold-atom vapors.
This function controls the quantum statistics of light emission from dense, inhomogeneous clouds of cold atoms of various spatial dimensionalities.
The results yield distinctive scaling behaviors on the microsecond timescale, including generalized exponential decay.
A detailed comparison is presented with a recent experiment on a cigar-shaped atomic ensemble [Y.~Dudin and A.~Kuzmich, {\it Science\/} {\bf 336\/}, 887 (2012)], in which Rb atoms are excited to a set of Rydberg levels.
\end{abstract}

\maketitle
Dipole-dipole interactions between atoms excited to Rydberg levels are very strong compared to interactions between atoms in their ground states~\cite{saffman_review}.
A novel application, proposed in Ref.~\cite{duan2000}, is that the Rydberg interaction could be utilized to entangle atoms, even separated by up to 10 microns, by controlling the atomic phase shifts accumulated in the presence of laser excitation.
As a consequence Rydberg interactions offer great promise --- e.g., in the areas of ultracold atomic physics and quantum information physics --- for the generation and control of many-body quantum states~\cite{bloch_manybody}, sources of nonclassical light emission and the realization of fast quantum gates~\cite{saffman_molmer_2009,*lesanovskyPRA2010,*pfau2011,*Gorshkov,*pritchard12}.

The Rydberg blockade is a paradigm for strong Rydberg atom interactions: an atom, once excited to a Rydberg level, induces a level shift of nearby atoms that prevents their excitation~\cite{lukin2001,*saffman2002,*saffmannature,*grangiernature}.
This process promises near-perfect efficiency for single photon emission and realization of quantum gates.
For the blockade to be effective, the interaction strength between every atom pair should exceed the Rabi frequency and linewidth of the excitation laser; this requirement imposes stringent limits on the size of the atomic cloud.

A different interaction regime may be reached by loading an atomic ensemble with a Poissonian number distribution of laser-induced Rydberg excitations~\cite{barianishort}.
Interactions then cause temporal dephasing of states with more than one excitation: an effect based on the variation of the interatomic potential in a spatially extended ensemble.
The retrieval process maps the spin wave pair correlations onto the emitted light,
via a phase matching condition~\cite{barianilong,*grangierretrieval}, promising a fast, high-quality, single-photon source.
The efficiency of the source is limited by the probability amplitude of the single-excitation component of the many-body wavefunction.
The experimental observation of spin wave dephasing has recently been reported~\cite{Dudin2012}.
That the interaction mechanism is dominated by dephasing, as opposed to Rydberg blockade, was inferred from the evolution of the excited fraction of Rydberg atoms as a function of the Rabi frequency and by the absence of many-body oscillations,
typical of the blockade regime.

In this Paper, motivated by the desire to obtain a closed expression for the time-dependence of the spin wave pair correlation function, we present a theory of the spin wave dynamics based on Rydberg atom short- and long-ranged interactions.
By using asymptotic methods appropriate for the long-time regime, we obtain analytical expressions for the dynamics of the spin wave pair correlation function, which controls the sub-Poissonian quantum statistics of the emitted light field and the speed of the single-photon protocol.
In order to be able to address highly asymmetric clouds, we derive expressions for samples of varying spatial dimensionality $D$.
We present results for both van der Waals (vdW) and dipole-dipole interaction potentials, and show that experimental data reported in Ref.~\cite{Dudin2012} is in agreement with the vdW theory.
\def\ark{r}

\noindent\emph{Spin wave dephasing}.
We consider a gas of $N$ atoms of fixed positions $\vv{r}_{\mu}$ (with $\mu=1,\ldots,N$),
laser excited from the atomic ground state $\ket{g}$ to a Rydberg level $\ket{\ark}$.
We denote by $\ket{G}$ the $N$-atom ground state,
and we denote by
$\ket{\mu \nu}\equiv\ket{g_1,\ldots,\ark_{\mu},\ldots,\ark_{\nu},\ldots,g_N}$
the state in which atoms $\mu$ and $\nu$
(and only those atoms) are excited from $\ket{g}$ to $\ket{r}$.
In the limit in which only a few atoms are excited,
we can effectively describe collective excitations by means of quasi-bosonic spin waves
associated with the destruction operator
$\hat{S}_{\vv{k}_0} =
(1/\sqrt{N})
\sum_{\mu = 1}^{N}{\rm e}^{-i\vv{k}_0\cdot\vv{r}_{\mu}}
\pro{g_{\mu}}{r_{\mu}}$,
where $\vv{k}_0$ is the wavevector of the spin wave, which is fixed by the excitation process.
We assume that the laser excitation prepares the system at time $t=0$ in the state
\begin{equation}
\ket{\Psi(t)}|_{t=0}=
\sum_{m = 0}^{\infty}
\frac{c_{m}}{\sqrt{m!}}
\left(\hat{S}^{\dagger}_{\vv{k}_0}\right)^{m}
\ket{G}.
\label{eq:Poisson}
\end{equation}
As the excitation is assumed to be weak, the distribution $|c_m|^2$ is typically a non-increasing function of $m$.
Two-body interactions between Rydberg-excited atoms $\mu$ and $\nu$ give a level-shift
$\Delta_{\mu \nu} = V_{\mu\nu}/\hbar$
that depends on the specific form of the interaction potential $V_{\mu\nu}$.
Here, we shall assume that
$V_{\mu\nu} = C_{\alpha}/|\vv{r}_{\mu} - \vv{r}_{\nu}|^{\alpha}$,
with $\alpha = 3$ or $6$ and $C_{\alpha}$ determined by the Rydberg target state~\cite{saffman_review,saffman08,kaulakys}.
After excitation, the $N$ atoms interact for a time $t$.
As a result the atomic states are phase shifted; e.g., in the two-body sector,
$\ket{\mu \nu} \rightarrow \exp(-i \Delta_{\mu\nu} t) \ket{\mu \nu}$.
More generally, the phase shift for a state with $m$ excitations, $\ket{\mu_1 \ldots \mu_m}$,  is given by
$\phi_{\mu_{1}\ldots\mu_{m}} =
 - t \sum_{1\le j<i\le m} \Delta_{\mu_i\mu_j} $.
The spin-wave pair correlation function, after storage time $t$,
$g^{(2)}(t)\equiv
\langle
\hat{S}_{\vv{k}_0}^{\dagger}(t)
\hat{S}_{\vv{k}_0}^{\dagger}(t)
\hat{S}_{\vv{k}_0}^{\phantom{\dagger}}(t)
\hat{S}_{\vv{k}_0}^{\phantom{\dagger}}(t)\rangle /
\langle
\hat{S}_{\vv{k}_0}^{\dagger}(t)
\hat{S}_{\vv{k}_0}^{\phantom{\dagger}}(t)\rangle^2$,
is given, with $j_{m-k}\equiv m!/[(m-k)!N^{m+k}],$ by
\begin{equation}
g^{(2)}(t)=
\frac{\displaystyle\sum_{m \geqslant 2}
|c_m|^2 j_{m-2}
\displaystyle\sum_{\mu_1..\mu_{m-2}}
\left|\displaystyle\sum_{\nu_1,\nu_2}
{\rm e}^{i \phi_{\mu_1..\mu_{m-2} \nu_1 \nu_2}} \right|^2}
{\left[\displaystyle\sum_{m \geqslant 1} |c_m|^2 j_{m-1}
\!\!\!
\displaystyle\sum_{\mu_1..\mu_{m-1}}
\left|\sum_{\nu}{\rm e}^{i \phi_{\mu_1..\mu_{m-1} \nu}}\right|^2\right]^2},
\label{eq:g2dephasing}
\end{equation}
where we have assumed  that $N \gg 1$. The term for $m=1$ in the denominator corresponds to $|c_1|^2$, as states having a single excitation are not phase shifted.
For $m  = 2$ in the numerator, we have only the sum within the absolute value that we define as the atom pair interference function
\begin{equation}
\mathcal{P}(t)\equiv
\frac{1}{N^2} \sum_{\mu =1}^{N} \sum_{\nu \neq \mu}{\rm e}^{- i\Delta_{\mu\nu} t}.
\end{equation}
It is known that $g^{(2)} (0)= 1$ for a Poisson distribution of $c_m$.
Interactions cause destructive interference of the distinct atom pair contributions, resulting in temporal decay of $\mathcal{P}(t)$.
Higher-order interference contributions may be expressed in terms of the two-body result
\begin{equation}
\sum_{\mu_1..\mu_{m-2}} \left|\sum_{\nu_1,\nu_2} {\rm e}^{i \phi_{\mu_1...\mu_{m-2} \nu_1 \nu_2}}\right|^2 \rightarrow N^{m+2} \left| \mathcal{P}(t) \right|^{m(m-1)},
\label{eq:multiple}
\end{equation}
where we use the results of Ref.~\cite{grangiernonclassical}.
For long enough times, $t \gg t_c$ where $t_c$ is a characteristic time associated with the specific interaction potential,
$|\mathcal{P}(t)| \ll 1$ and, therefore, the correlation function reduces to the form~\cite{barianishort}
\begin{equation}
g^{(2)}(t)
\xrightarrow{t \gg t_c}
\frac{\abs{c_2}^2  2 \abs{\mathcal{P}(t)}^2}
{ \left[ \abs{c_1}^2 + \abs{c_2}^2  2 \abs{\mathcal{P}(t)}^2 \right]^2}.
\label{eq:g2_Ptilde}
\end{equation}
In Ref.~\cite{barianilong}, we exhibit the direct relationship between states of spin wave excitation
and those of photons emitted into a phase-matched mode of the field. As a consequence, the spin wave pair correlation function is equal to the Glauber's normalized second-order correlation function of the phase-matched field mode.

\begin{figure}[tbp]
\begin{center}
\includegraphics[width =  \columnwidth]{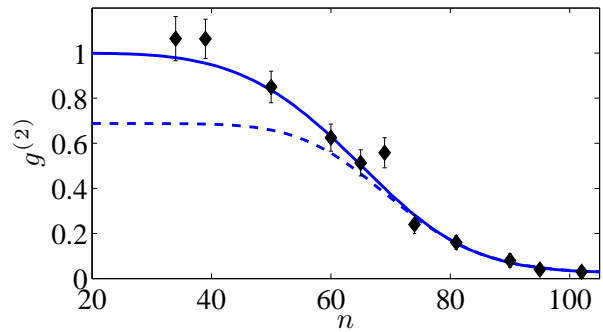}
\caption{(Color online). Behavior of the spin wave pair correlation function $g^{(2)}$ vs.~$n$
for an asymmetrical cloud having $(w_z,w_{\perp}) = (15,6.4)\,\mu{\rm m}$, compared with the experimental data from Ref.~\cite{Dudin2012} (Black dots). Also shown are results from the double spin wave model (Blue, dashed line). Experimental data are reprinted with permission from AAAS and authors.}
\label{fig:dephasing_vs_n}
\end{center}
\end{figure}

\noindent\emph{Dephasing in a cigar-shaped ensemble}.
In this section, by using formula~\eq{g2dephasing}, we numerically calculate $g^{(2)}(t)$ for a cigar-shaped ensemble and compare the results with the data of the experiment described in Ref.~\cite{Dudin2012}. In that work, the spin wave correlations were inferred by measuring the probability of two coincident photoelectric detection events, in a phase matched mode determined by the experimental setup. Instead of studying the temporal dynamics for a given Rydberg level,
the principal quantum number $n$ was varied for a fixed storage-time $T_s = 0.3\,\mu{\rm s}$.
As the phase shifts are given by $V_{\mu\nu}T_s/\hbar$,
and the Rydberg interaction scales with $n$,
the measurements effectively map out the dynamics.
The longitudinal size of the ensemble is set by the optical lattice beam, of waist $w_z = 15 \mu{\rm m}$;
the transverse area is determined by the overlap of two laser beams, each of waist $9\mu{\rm m}$, giving a transverse waist of the excitation zone $w_{\perp}$ of $(9/\sqrt{2})\,\mu{\rm m}$.
For an ensemble of peak density $\rho_0 \sim 10^{12}\,{\rm cm}^{-3}$,
the number of interacting atoms is given by
$N=(\pi/2)^{3/2}\rho_0\,w_z\,w_{\perp}^2\sim 1200$.
In the present work, we determine the interatomic potentials for different $n$ via a single-channel model,
$\ket{ns_{1/2},ns_{1/2}}\leftrightarrow\ket{np_{3/2},(n-1)p_{3/2}}$,
which is based on a semiclassical calculation of the dipole transition matrix elements~\cite{saffman08,*kaulakys}.
We evaluate the atom pair interference function by Monte Carlo integration, and use it to determine the contribution of multiple excitations: the result converges once we have included of the order of $10$ excitations.
We sample $N=1000$ atoms, although the result is essentially independent of $N$, provided $N$ is much larger than the number of excitations.
We assume that the initial spin-wave distribution state is Poissonian \eq{Poisson} with unit mean, i.e.,
$c_m = 1/\sqrt{e\, m!}$. To facilitate comparison with the data of Ref.~\cite{Dudin2012}, the theoretical values are adjusted to account for the measured photodetection background noise $g^{(2)}_{bg}$, $g^{(2)} \rightarrow (1 - g^{(2)}_{bg})g^{(2)} + g^{(2)}_{bg}$.
Figure~\ref{fig:dephasing_vs_n} shows striking agreement between the model and the experimental data over the full range of principal quantum numbers measured.  We note that a model retaining only single and double excitations is good for large principal quantum numbers, i.e., $n \gtrsim 60$.

\noindent\emph{Analytical treatment of spin wave dephasing}.
To develop an analytical theory of the spin wave dynamics
in the limit $N \gg 1$, we define $\mathcal{P}(t)$ via
\begin{equation}
\mathcal{P}(t)\equiv
\int d\Delta\,{\rm e}^{ - i\Delta t}\,
p(\Delta),
\label{eq:phase_distr}
\end{equation}
where $p(\Delta)$ is the probability distribution for the interaction-induced frequency shift $\Delta$
associated with the ensemble of atom locations.
We note that $p(\Delta)$ is determined by the interaction potential and by the geometry
of the atomic ensemble.

To determine $p$,
we consider a $D$-dimensional
isotropic,
ensemble having a Gaussian density profile;
the formalism makes it possible to deal qualitatively also with anisotropic samples
whose aspect ratio can be experimentally engineered via a combination of trap and excitation beams~\cite{bloch_manybody}.
The two-particle spatial distribution $P$ is given by
$ P_D(\vv{r}_1,\vv{r}_2) = \mathcal{N}_D\,{\rm e}^{- (r^2_1 + r^2_2)/(2\sigma^2)}$,
where $w = 2 \sigma$ is the waist of the cloud,
and the normalization $\mathcal{N}_D$ is given by $\mathcal{N}_D = (2\pi \sigma^2)^{-D}$.
Then the probability density for finding two particles separated by the {\it vector\/} $\vv{R}$ is given by
\begin{align}
P_D(\vv{R}) &=
\int
d^{D}{r}_1\,
d^{D}{r}_2\,
P(\vv{r}_1,\vv{r}_2)\,
\delta(\vv{R}-\vv{r}_{1} + \vv{r}_{2}) \nonumber \\
&= \frac{{\rm e}^{- R^2/(4\sigma^2)}}{(4\pi\sigma^2)^{D/2}}.
\end{align}
Hence, one has that the distribution $P_D(R)$ of interparticle {\it separations\/} $R\equiv\vert\vv{R}\vert$
in dimension $D$ is given by
$P_D(R)= S_{D-1} e^{- R^2/(4\sigma^2)}/(4\pi\sigma^2)^{D/2}$,
where $S_{D-1}$ [$=\left(2\pi^{D/2}/\Gamma(D/2)\right)R^{D-1}$]
is the $(D-1)$-dimensional area of the surface of a $D$-dimensional ball of radius $R$.
From this distribution, we have access to the probability distribution of a function $f(R)$, via the standard Jacobian transformation:
\begin{equation}
p_D\big(f(R)\big) = \frac{1}{{\vert{df/dR}\vert}} \frac{S_{D-1} e^{- R^2/(4\sigma^2)}}{(4\pi\sigma^2)^{D/2}} .
\label{eq:jacobian}
\end{equation}
We use this result to determine the probability distribution of Rydberg atom level shifts for different interaction potentials, enabling calculation of the atom pair interference function via asymptotic methods.

\noindent\emph{\lq\lq Isotropic\rq\rq\ dipole-dipole interaction}.
The interaction between states of nonzero dipole moment is proportional to $C_3/R^3$, where $C_3$ is determined by the contributing scattering channels~\cite{kaulakys}.  One way to achieve this interaction is by coupling
near-degenerate atomic orbitals of different parity via an electrostatic field,
so that the atoms acquire a dipole moment aligned along the field~\cite{pfauDCfield}.
For systems of reduced dimensionality (i.e., $D$ being 1 or 2), the interaction can be made isotropic by applying the external field in a direction orthogonal to the system; for the case of $D = 3$, isotropy is not truly achievable, so an isotropic theory should provide only a qualitative description.

\begin{figure}[tbp]
\begin{center}
\includegraphics[width =  \columnwidth]{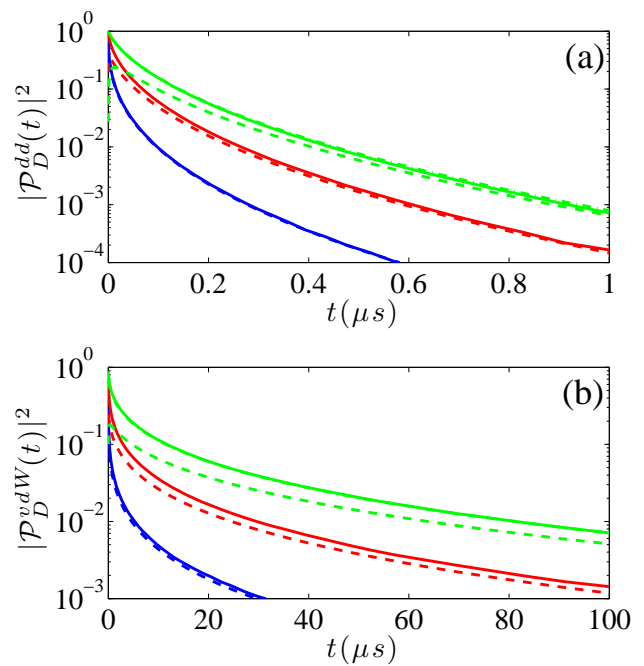}
\caption{\label{fig:deph_longT}(Color online).
Asymptotic behavior of atom pair interference function
for double excitations of the $100s$ level.
(a)~Isotropic dipole-dipole interactions:
solid lines are numerical simulation for $D=1$ [blue (black in BW)], $2$ [red (dark gray)] and $3$ [green (light gray)],
compared with the long-time asymptotic behavior given by the modulus square of Eq.~\eq{P_dd} (dashed lines).
The green (light gray) dot-dashed line includes the first asymptotic correction for $D=3$.
(b)~vdW interactions: (color legend as in Fig.~a)
the long-time behavior shows the square modulus of formula~\eq{vdW_Ptilde}.
The first-order asymptotic correction for $D=3$ is indistinguishable from the solid line.
Parameters are given in the text.}
\end{center}
\end{figure}

We now calculate the atom pair interference function: $\mathcal{P}^{dd}_{D} (\tau) = \int_{0}^{\infty} d\kappa\,p^{dd}_D(\kappa)\,{\rm e}^{- i \kappa \Omega_{dd} t}$. Here, we define the dimensionless time $\tau\equiv\Omega_{dd}t$, where $ \hbar \Omega_{dd} \equiv (C_3/\sigma^3)$, and the normalized interaction energy $\kappa\equiv(\sigma/R)^3$, whose probability distribution
is then given by Eq.~\eq{jacobian},
\begin{equation}
p^{dd}_{D}(\kappa)=\frac{1}{2^{D-1}\,3\,
\Gamma(D/2)} \frac{1}{\kappa^{1+D/3}}{\rm e}^{-{1}/{4 \kappa^{2/3}}}.
\label{eq:Pofk}
\end{equation}
By using the method of steepest descents (see, e.g., Ref.~\cite{bender}),
we evaluate the long-time asymptotic behavior of the integral, obtaining at leading order
\begin{equation}
\mathcal{P}^{dd}_{D}(\tau)\xrightarrow{\tau\gg 1}
\mathcal{A}_{D}\,
\tau^{{(D-1)}/{5}}
{\rm e}^{-\mathcal{B}\tau^{2/5}},
\label{eq:P_dd}
\end{equation}
where
$\mathcal{B}\equiv(5/2)\times 6^{-3/5}\,{\rm e}^{-i{\pi}/{5}}$, and $\mathcal{A}_{D}\equiv
\sqrt{\pi/5} \times {\rm e}^{-i\pi(D-1)/10}
6^{(D-1)/5}/\big(2^{D-2}\,\Gamma(D/2)\big)$.
According to Eq.~(\ref{eq:g2_Ptilde}), the asymptotic decay of $g^{(2)}$ follows the square modulus of $\mathcal{P}^{dd}_D(\tau)$, which is given by
$ \abs{\mathcal{P}^{dd}_{D}(\tau)}^2
\xrightarrow{\tau \gg 1}
\abs{\mathcal{A}_D}^{2}\,
\tau^{\frac{2}{5}(D-1)}\,
{\rm e}^{- 2\real{\mathcal{B}}\,\tau^{2/5}}.$
We note that the next asymptotic correction to $\mathcal{P}^{dd}_{D} (\tau)$, multiplies the latter by the complex factor
$ \{
1 +
{\rm e}^{i\frac{\pi}{5}}/(6\tau)^{2/5}
[\frac{3}{5} (1+\frac{D}{3})(D-2) + \frac{14}{15}]
\}$.
In \fig{deph_longT}a, discussed further below, we compare the asymptotics with numerical results generated for an ensemble of atoms at randomly distributed positions \footnote{See Supplemental Material for short-time behavior of $\mathcal{P}^{dd}_{3} (\tau)$}.

\noindent
\emph{Van der Waals interaction}. In the absence of external electromagnetic fields, the interaction between atoms excited to Rydberg levels has a dipole-dipole character at small separation and crosses over to the vdW (i.e., $R^{-6}$) form at larger separations~\cite{saffman08}.
As the typical size of the ensembles under consideration is much larger than the cross-over radius, we assume a vdW interaction over the whole cloud.
In analogy with the discussion of the dipole-dipole case, we define a dimensionless vdW energy $\varkappa\equiv(\sigma/R)^6$, distributed according to
\begin{equation}
p^{vdW}_{D}(\varkappa)=
\frac{1}{2^{D}\,3\,\Gamma(D/2)}
\frac{1}{\varkappa^{1+D/6}}{\rm e}^{-{1}/{4 \varkappa^{1/3}}}.
\label{eq:Pofk}
\end{equation}
An asymptotic expression for the atom pair interference function~\eq{phase_distr} is found using the method of steepest descents as before,
\begin{equation}
\mathcal{P}^{vdW}_{D} (\tau) \xrightarrow{\tau \gg 1}
\mathcal{C}_D\,\tau^{\frac{D-1}{8}}\,
{\rm e}^{- \mathcal{F} \tau^{1/4}},
\label{eq:vdW_Ptilde}
\end{equation}
where the dimensionless time is $\tau\equiv  \Omega_{vdW} t$, $\hbar \Omega_{vdW} \equiv (C_6/\sigma^6)$, and we define the coefficients $\mathcal{C}_D\equiv
\sqrt{2 \pi}\,
{\rm e}^{-i\frac{\pi}{16}(D-1)}\,
12^{(D-1)/8}/\big[2^D \Gamma(D/2)\big]$
and
$\mathcal{F}\equiv
\big(12^{1/4}/3\big)\,
{\rm e}^{-i{\pi}/{8}}$.
We note that the next order asymptotic correction maps
$\mathcal{P}^{vdW}_{D}(\tau)
 \rightarrow
 \mathcal{P}^{vdW}_{D}(\tau)\times
 \{1 +
 {\rm e}^{i{\pi}/{8}}/(12\tau)^{1/4}
 \big[\frac{3}{2} (1+\frac{D}{6})(\frac{3D}{4}-1) + \frac{35}{24}\big]
 \}$.
In \fig{deph_longT}b, we show the dephasing dynamics for the vdW interaction for atomic ensembles of various dimensionality but common waist size.

These results may be compared with those for dipole-dipole interactions, shown in \fig{deph_longT}a.
We compare the asymptotics with numerical calculations for the target Rydberg level $100s$, for which
$C_3/\hbar=2\pi\times 10^{5}\,{\rm MHz}\,\mu{\rm m}^3$
and
$C_6/\hbar=2\pi\times 5.3\times 10^7\,{\rm MHz}\,\mu{\rm m}^6$.
For an ensemble of waist $w=30\,\mu{\rm m}$, we obtain typical level shifts of
$\Omega_{dd}=2\pi\times 30\,{\rm MHz}$
and
$\Omega_{vdW}=2\pi\times 4.7\,{\rm MHz}$.
We note that for the vdW case, decay is roughly two orders of magnitude slower than for the dipole-dipole interaction.
This result is dependent on the size of the ensemble, which sets the typical interaction strengths $(\Omega_{dd} > \Omega_{vdW})$, and on the range of the interaction, which determines the power-law in the generalized exponential decay $(2/5 > 1/4)$. The longer range of the dipole-dipole interaction makes it more effective for fast dephasing in a mesoscopic ensemble \footnote{Dipole-dipole interactions may also be induced by resonant microwave driving of Rydberg levels. This case is discussed in the Supplemental Material.}.
In \fig{deph_longT} we also observe that the leading-order asymptotics
is more accurate for $D=1$ and $2$ than for $D=3$.
The results for $D=3$ are significantly improved by including the next asymptotic correction.
The analytical and numerical results are then almost indistinguishable on the scale shown.
Although the cigar-shaped ensemble does not have a Gaussian density profile, as assumed in our asymptotic analysis, it is possible to qualitatively model the data of \fig{dephasing_vs_n} for $n \gtrsim 60$, if we take the dimension of the gas to be nonintegral, in fact $D = 1.3$.
We remark that we are showing the behavior of $\mathcal{P}(\tau)$ because we are interested in the limit $\mathcal{P}(\tau) \ll 1$, where $g^{(2)}$ is dominated by the two-body interactions. Contributions to the spin wave pair correlation function~\eq{g2dephasing} from more than two excitations can be assessed via formula~\eq{multiple}.

In conclusion, we have given results for the long-time asymptotic dynamics of Rydberg atom spin waves that have been prepared via laser excitation of a cold atomic ensemble.
Our results provide analytical scaling behaviors that are of value in assessing, e.g., the limits in speed-up achievable by using Rydberg atom interactions for single-photon protocols relevant to quantum information processing~\cite{yarik_phd}.
For clouds having a Gaussian density profile, the decay follows a generalized exponential in time, with a rational power in the exponent that is governed by the range of the interaction potential.
Numerical results for a cigar-shaped cloud are in good agreement with a recent experimental measurement of spin wave dephasing in atomic Rb~\cite{Dudin2012}.

We thank Y.O.~Dudin and A.~Kuzmich for discussions about their experiment.
F.B.\ and T.A.B.K.~acknowledge support from the Air Force Office of Scientific
Research Atomic Physics Program, Quantum Memories
Multidisciplinary University Research Initiative, and NSF.
P.M.G.~acknowledges support from NSF via award No.~DMR 09 06780 and DMR 12 07026.

\begin{widetext}

\section{Supplemental Material: Short-time behavior of the atom pair interference function for dipole-dipole interaction for D = 3}

We have calculated the short-time dynamics for $D=3$ of the atom pair interference function for the case of isotropic dipole-dipole interaction. By employing the method of matched asymptotic expansions \cite{bender}, we have obtained the result
\begin{equation}
\mathcal{P}^{dd}_3 (\tau) \xrightarrow{\tau \rightarrow 0} 1 - \frac{\sqrt{\pi}\tau}{12} \left[1 - \frac{6i}{\pi} \left( \mathrm{ln}\frac{\tau^{1/3}}{2} + \frac{5}{6} \gamma   - \frac{1}{3} \right) \right],
\end{equation}
where $\gamma = 0.5772\ldots$ is the Euler-Mascheroni constant.

\begin{figure}[htbp]
\begin{center}
\includegraphics[width = \textwidth]{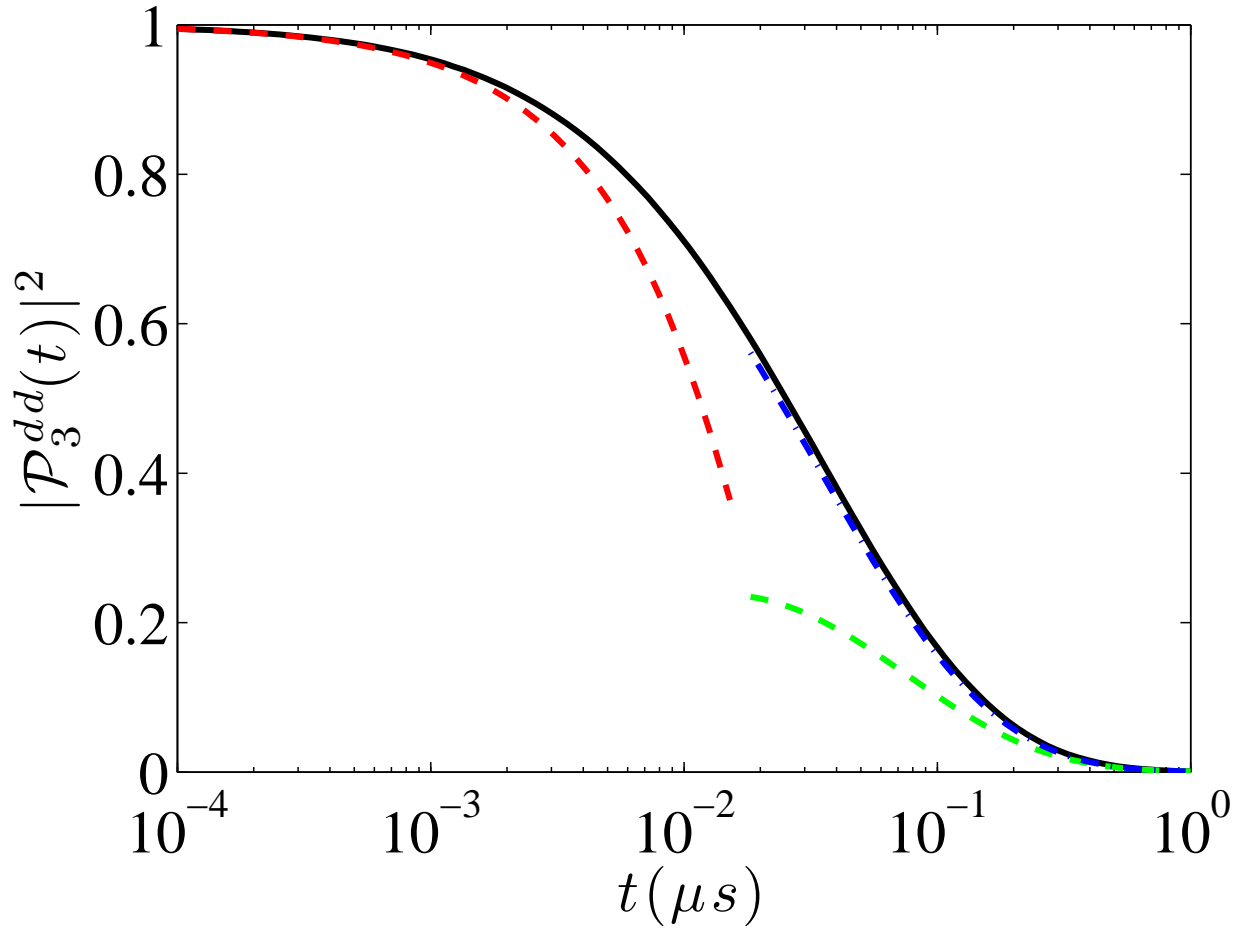}
\caption{(Color online). Comparison of short-time [Red dashed line from Eq.(13)] and long-time (Green dashed line from the main text of the paper) leading asymptotics of $\abs{\mathcal{P}_3(t)}^2$ with the numerical results (Black solid line). The Blue dot-dashed line includes the first order long-time asymptotic correction. 
We consider the target Rydberg level $100s$ for Rb, and an ensemble waist, $w = 30 \mu$m.}
\label{fig:dip_dip_3D}
\end{center}
\end{figure}

\section{Supplemental Material: Microwave driven dipole-dipole interaction}
We have shown in the main text of the paper that, for a mesoscopic atomic ensemble, dipole-dipole interactions cause the rapid decay of the spin wave pair correlation function.
Achieving the necessary stability to produce this interaction with static electic fields is, however, experimentally challenging~\cite{saffman_review}.
An alternative approach involves driving a transition between neighboring Rydberg levels by means of microwave pulses: the mixing of opposite-parity Rydberg orbitals provides the atoms with an oscillating dipole moment~\cite{adamsmuwave}. 
This technique has recently found interesting applications in experiments \cite{adams2012,walker2012}. 
Reference~\cite{barianishort} elaborates a scheme in which induced dipole-dipole interactions in an atomic ensemble may be used to speed up the production of single photons, by orders of magnitude, over a non-interacting atomic ensemble. 
The single photon protocol involves successive Ramsey pulse sequences - each of which implies two microwave pulses separated by an interval during which the Rydberg atoms interact.
A simple analysis of a single Ramsey sequence shows that the amplitude picked up by a double excitation due to interactions is given by: $\ket{\mu\nu} \rightarrow {\rm e}^{-i \Delta_{\mu\nu} t} \cos(\Delta_{\mu\nu} t) \ket{\mu\nu}$. This result differs from the case of dipole-dipole and van der Waals interaction, where only a phase shift is introduced by the interaction.
The pair correlation function at large time may still be calculated from Eq.~(5) of the main text, but with the replacement
$\mathcal{P}(t)\rightarrow\frac{1}{2}[1+\mathcal{P}(2 t)]$. 
This change means that the correlation function relaxes to a finite value smaller than unity (sub-Poissonian statistics) after a single pulse; in order to generate true single photons, multiple repetition of the Ramsey sequence are required.
In \fig{dip_res} we compare the analytical result (lines) for the asymptotic behavior of $g^{(2)}$ after a single Ramsey cycle with simulations (dots) based on the proposal contained in Ref.~\cite{barianishort}. The fast initial decay of $g^{(2)}$ is of particular interest, as it determines the limiting speed of single photon production.
The present theory suggests that this decay is a generalized exponential, of the form
$[\tau^{2/5}\,{\rm e}^{-\real{\mathcal{B}}(2\tau)^{2/5}}].$
Although this result
is obtained using an isotropic interaction, it is still able to reproduce the main features of the numerical calculation of $g^{(2)},$ i.e., the steep initial decay to a minimum and recovery to a constant asymptotic value.
Overall, the qualitative agreement seems good.
The discrepancy at long times is principally due to the anisotropy of the dipole-dipole interaction, which is included in the numerical simulations.

\begin{figure}[htbp]
\begin{center}
\includegraphics[width =  \columnwidth]{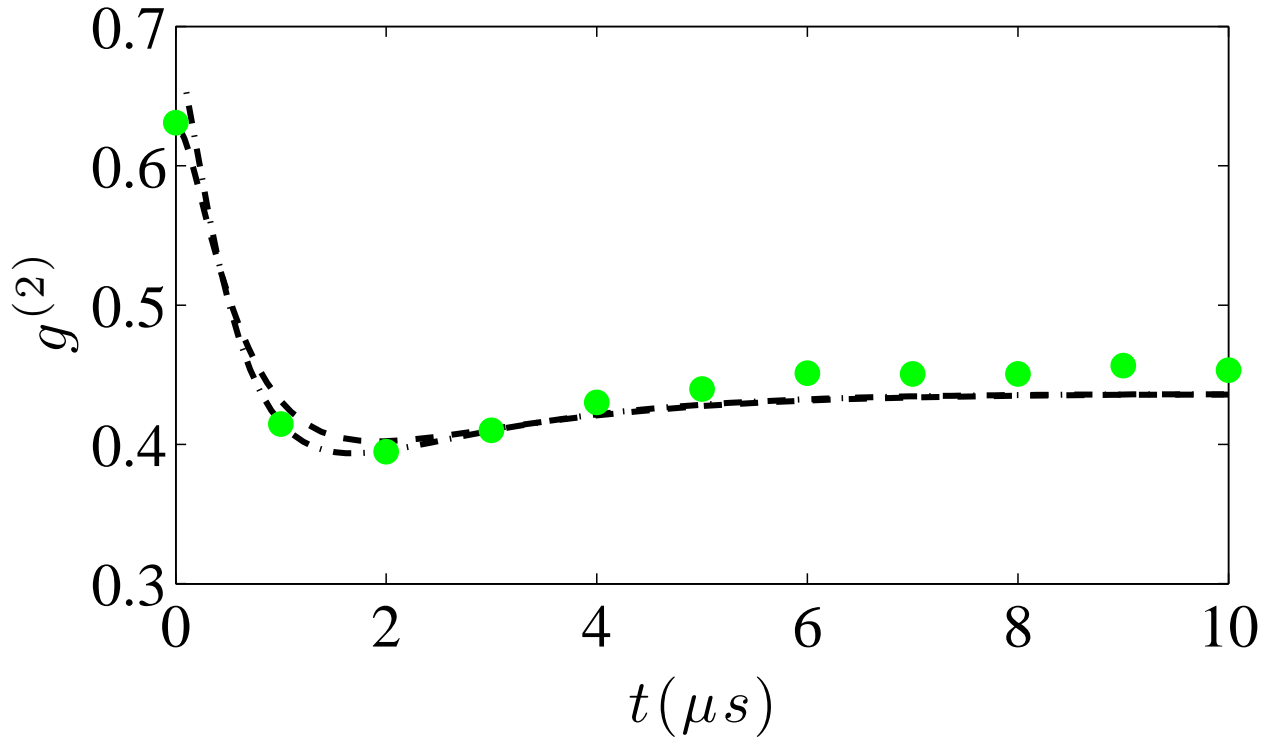}
\caption{(Color online) Comparison between (i)~the asymptotic behavior of correlations for microwave-induced dipole-dipole interactions in a three-dimensional sample having $w = 60\,\mu{\rm m}$ and (ii)~numerical analysis of the full anisotropic interaction~\cite{barianishort} (green dots).  The correlation function is calculated from the two-body integral using Eq.~(5) in the main text.
Dashed line: leading-order asymptotics;
dot-dashed line: including the first asymptotic correction.}
\label{fig:dip_res}
\end{center}
\end{figure}

\end{widetext}

\bibliography{references_analytical_dephasing}
\end{document}